\documentstyle[12pt]{article}

\font\titlefont=cmbx10 scaled \magstep3

\begin{document}
\input{epsf}

\begin{flushright}
quant-ph/9704035  \\ 
April 17, 1997 \\
TUPT-97-5
\vspace*{1cm}
\end{flushright}

\begin{center}
{\titlefont Electromagnetic Vacuum Fluctuations \\
\vskip 0.2in
and Electron Coherence II: \\
\vskip 0.2in
Effects of Wavepacket Size }
\vskip .4in
L.H. Ford\footnote{email: ford@cosmos2.phy.tufts.edu} \\
\vskip .2in
Institute of Cosmology\\
Department of Physics and Astronomy\\
Tufts University\\
Medford, Massachusetts 02155\\
\end{center}

\vskip .3in

\begin{abstract}
If one analyzes the effects of electromagnetic vacuum fluctuations upon
an electron interference pattern in an approximation in which the electrons
follow classical trajectories, an ultraviolet divergence results. It is
shown that this divergence is an artifact of the classical trajectory 
approximation, and is absent when the finite sizes of electron wavepackets
are accounted for. It is shown that the vacuum fluctuation effect has  a 
logarithmic dependence upon
the wavepacket size. However, at least in one model geometry, this dependence
cancels when one includes both vacuum fluctuation and photon emission effects.
\end{abstract}
\vspace{0.6cm}
PACS categories: 03.75.-b,  12.20.-m
\newpage

\baselineskip =14pt

\section{Introduction}

     In a previous paper \cite{F93}, henceforth I, the effects of 
the quantized electromagnetic field upon coherent electrons were
investigated. It was found that the electron interference patterns are
modified by vacuum fluctuation and photon emission effects. If $\psi_1$
is the amplitude for an electron to travel between two spacetime points
along path $C_1$ and $\psi_2$ is that to travel between the same pair of
points along path $C_2$, then the familiar result of elementary quantum
mechanics for the number density of electrons at the endpoint is
\begin{equation}
n_0 = |\psi|^2 = |\psi_1 +\psi_2|^2 = 
|\psi_1|^2 +|\psi_2|^2 +2 Re (\psi_1 {\psi_2}^*)\,,
                                          \label{eq:numdens0}
\end{equation}
However, this result does not take account of the coupling of the electrons
to the quantized electromagnetic field. In the presence of such coupling,
it was shown in I that the number density now becomes
\begin{equation}
n = |\psi_1|^2 +|\psi_2|^2 +
                2\,e^W Re (e^{i\phi}\psi_1 {\psi_2}^*)\, . \label{eq:numdens1}
\end{equation}
where $\phi$ is a phase shift which may be taken to include both the
Aharonov-Bohm phase due to any classical electromagnetic fields, as well
as terms due to quantum electromagnetic effects, which are discussed in I.

In the present paper, we will be primarily concerned with the change in the
amplitude of the interference oscillations due to the $e^W$ factor. In an
approximation in which one assumes that the electrons move upon classical
worldlines, 
\begin{equation}
W = -2\pi\alpha \oint_{C}d{x_\mu}\oint_{C}d{{x'}_\nu} D^{\mu\nu}(x,x'),
      \label{eq:defw}
\end{equation}
where $\alpha$ is the fine structure constant, and $C = C_1 - C_2$ is the 
closed spacetime path obtained by traversing $C_1$
in the forward direction and $C_2$ in the backward direction \cite{units}. 
The photon Hadamard (anticommutator) function, $D^{\mu\nu}(x,x')$, is defined by
\begin{equation}
D^{\mu\nu}(x,x') =  {1 \over 2} \langle 0| \{ A^\mu (x), A^\nu (x') \}
                     |0 \rangle \,. \label{eq:twopoint}
\end{equation}
Equation~(\ref{eq:defw}) may be converted into a double surface integral
over the two dimensional timelike surface bounded by $C$ :
\begin{equation}
W = -2\pi\alpha \int d{a_{\mu\nu}}\int d{{a'}_{\rho\sigma}} 
      D^{\mu\nu;\rho\sigma}(x,x'),  \label{eq:surfaceint}
\end{equation}
where $d{a_{\mu\nu}}$ is the area element of this surface  and
\begin{equation}
D^{\mu\nu;\rho\sigma}(x,x') = {1 \over 2} \langle 0| \{ F^{\mu\nu} (x), 
                     F^{\rho\sigma} (x') \} |0 \rangle  
\end{equation}
is the Hadamard function for the field strengths.
Equation~(\ref{eq:surfaceint}) has the interpretation that the 
electrons are sensitive to vacuum fluctuations in regions from which 
they are excluded. This is analogous to the situation in the Aharonov-Bohm 
effect \cite{AB},
where the phase shift can depend upon classical electromagnetic fields in 
regions which the electrons cannot penetrate. 

   If $W < 0$, so that $e^W < 1$,  we may interpret
$W$ as describing the decohering effects of the quantized electromagnetic 
field. This issue will be discussed further in Sect. \ref{sec:sum}.
 Equation~(\ref{eq:defw}) may be obtained by averaging over a
fluctuating Aharonov-Bohm phase. Let
\begin{equation}
\phi = e \oint_{C}d{x_\mu}\, A^\mu (x)
\end{equation}
be the Aharonov-Bohm phase difference between paths $C_1$ and $C_2$ 
in the presence of the vector potential $A^\mu (x)$. If $A^\mu (x)$
is a fluctuating quantum field then $\phi$ undergoes fluctuations, and
we may average the phase factor using the relation
\begin{equation}
\Bigl \langle{\rm e}^{i\phi} \Bigr \rangle = 
{\rm e}^{-\frac{1}{2} \langle \phi^2 \rangle} 
\end{equation}
and obtain Equation~(\ref{eq:defw}). Decoherence due to a fluctuating
Aharonov-Bohm phase was discussed by Stern {\it et al} \cite{AS}.

This decohering effect was shown in I to consist of a mixture of
photon emission and vacuum fluctuation contributions.
Equation~(\ref{eq:defw}) may be split into two pieces using the 
relation
\begin{equation}
\oint_{C} \oint_{C} = \int_{C_1} \int_{C_1} + \int_{C_2} \int_{C_2}
                      -\int_{C_1} \int_{C_2} -\int_{C_2} \int_{C_1} \,.
\end{equation}
The cross terms involving integrations over both $C_1$ and $C_2$ describe
the effects of photon emission, whereas the other two terms describe the
effects of the vacuum fluctuations. That is, we may write
\begin{equation}
W = W_V + W_\gamma \,,
\end{equation}
where
\begin{equation}
W_V = -2\pi\alpha \biggl(\int_{C_1}d{x_\mu} \int_{C_1}d{{x'}_\nu} + 
\int_{C_2}d{x_\mu} \int_{C_2}d{{x'}_\nu} \biggr) D^{\mu\nu}(x,x')
      \label{eq:defwv}
\end{equation}
describes the effects of the vacuum fluctuations and
\begin{equation}
W_\gamma = 2\pi\alpha \biggl(\int_{C_1}d{x_\mu} \int_{C_2}d{{x'}_\nu} + 
\int_{C_2}d{x_\mu} \int_{C_1}d{{x'}_\nu} \biggr) D^{\mu\nu}(x,x')
      \label{eq:defwg}
\end{equation}
describes those of photon emission.

However, $W_V$ as given by Eq.~(\ref{eq:defwv}) is infinite. Formally,
this arises from the singularity of the Hadamard function 
$D^{\mu\nu}(x,x')$ when $x = x'$. (This problem does not arise for 
$W_\gamma$ because the integrations in Eq.~(\ref{eq:defwg}) are over
two distinct paths.) Physically, the divergence of $W_V$ is due to 
use of an approximation in which the electrons travel upon classical
worldlines. In a more realistic treatment, which will be the primary
topic of this paper, the electrons are described by finite size 
wavepackets, and $W_V$ is finite. If one is primarily interested in
changes in $W_V$ due to the presence of boundaries, as was the case in
I and in Ref. \cite{F95}, then the classical trajectory approximation is
adequate. The infinite part of $W_V$ cancels when one takes the difference
of the situation with the boundary and without the boundary. 

\section{Finite Wavepackets}
 
We now wish to go beyond the classical trajectory approximation and 
incorporate the effects of the finite size of the electron wavepackets.
We will continue to assume that the spatial size of these wavepackets is
small compared to the separation between the paths $C_1$ and $C_2$, which
are now taken to be the average worldlines of the packets. Thus, the
probability density for an electron which takes path $C_1$ from the source 
to the detector is nonzero inside of a world tube centered around $C_1$.

As was discussed in I, $W_V$ may be obtained from the vacuum 
persistence amplitude \cite{Schwinger}, which is given by
\begin{equation}
\langle out|in \rangle = \exp \bigl[-{i\over 2} \int j_\mu (x)j_\nu (x')
D_F^{\mu \nu} (x,x') {d^4}x {d^4}x' \bigr]\,, \label{eq:vpa}
\end{equation}
where $D_F^{\mu \nu} (x,x')$ is the Feynman propagator, and $j_\mu (x)$
is the electric current density. We are here concerned with the magnitude,
but not the phase of this amplitude, so we may use the relation 
\begin{equation}
D_F^{\mu \nu} (x,x') = {1 \over 2} \bigl[ D_{ret}^{\mu \nu} (x,x')
 + D_{ret}^{\nu \mu} (x',x) \bigr] - iD^{\mu \nu} (x,x'),
\end{equation}
where ${D_{ret}}^{\mu \nu} (x,x')$ is the retarded Green's function, to write
\begin{equation}
|\langle out|in \rangle| = \exp \bigl[-{1\over 2} \int j_\mu (x)j_\nu (x')
D^{\mu \nu} (x,x') {d^4}x {d^4}x' \bigr]\,. \label{eq:mvpa}
\end{equation}
The general form of $W_V$ is now
\begin{eqnarray}
&&W_V = \ln \Bigl( |\langle out|in \rangle_1||\langle out|in \rangle_2|\Bigr) 
      =                                               \nonumber \\
&& \!\!\!\!\!\!\!   -\frac{1}{2} \left[
\int_{C_1} j_\mu (x)j_\nu (x')D^{\mu \nu} (x,x') {d^4}x {d^4}x' +
\int_{C_2} j_\mu (x)j_\nu (x')D^{\mu \nu} (x,x') {d^4}x {d^4}x' \right] \,, 
\end{eqnarray}
where the labels $1$ and $2$ refer to the paths $C_1$ and $C_2$, 
respectively. In general, $j^\mu$ should be the Dirac current for the
wavepacket state. However, we will here ignore magnetic moment effects,
so the Dirac current may be replaced by its convective part, and we will
assume nonrelativistic motion, so that the expression for the current
density in the Schr\"odinger theory may be used. In this case, 
$j^\mu = (j^0, {\bf j})$, where
\begin{equation}
j^0 = e |\psi|^2\,,
\end{equation}
and
\begin{equation}
{\bf j} = \frac{e}{2 i m} \bigl[ \psi^* {\bf \nabla}\psi - 
                         ({\bf \nabla}\psi^*) \psi \bigr] \,.
\end{equation}
Here $\psi$ is the Schr\"odinger wavefunction for the wavepacket state,
and $m$ and $e$ are the mass and charge of the electron, respectively.
To the extent that the wavepacket moves with a velocity ${\bf v}$ without
spreading, we have that
\begin{equation}
{\bf j} \approx  e\,{\bf v}\, |\psi|^2  \,.
\end{equation}
 In any case, in the nonrelativistic limit the dominant contribution to
$W_V$ comes solely from $j^0$. 

In the Feynman gauge, the Hadamard function takes the form
\begin{equation}
D^{\mu\nu}(x,x') =  \frac{\eta^{\mu \nu}}{4\pi^2 (x-x')^2}\,,
\end{equation}
where $\eta^{\mu \nu}$ is the Minkowski spacetime metric, which we take 
to have signature $(1,-1,-1,-1)$, and $(x-x')^2 = (t-t')^2 - 
|{\bf x} - {\bf x}'|^2$. Thus we can see that the contribution of 
the spatial components of the current are of order $v^2$, and can be 
neglected. Denote the probability density of the wavepacket by
$f({\bf x},t)$ :
 \begin{equation}
  f({\bf x},t) = |\psi|^2  \,.
\end{equation}
The contribution of the path $C_i$ to $W_V$ may now be expressed as
\begin{equation} 
W_V(C_i) = - \frac{\alpha}{2\pi} \int_{C_i} {d^4}x \,{d^4}x' \, 
             \frac{f(x) f(x')}{(x-x')^2} \,.
 \end{equation}
 This integral
is to be interpreted as a principal value. For the remainder of this 
paper, all integrals with poles in the range of integration are understood
to be principal values.

      We will restrict our attention to the case where spreading of
the wavepacket can be ignored. Let the trajectory of the center of the 
wavepacket traveling on path $C_i$ be given by ${\bf x}_i(t)$ and let
${\bf y} = {\bf x} - {\bf x}_i(t)$.  That is, ${\bf y}$ is the displacement
from the center of the wavepacket. In the absence of spreading,   
$f({\bf x},t) = f({\bf y})$, and we can write
\begin{equation} 
W_V(C_i) = - \frac{\alpha}{2\pi} \int dt \, dt' \int {d^3}{\bf y}\,{d^3}{\bf y'} 
   \; \frac{f({\bf y}) f({\bf y'})}
           {(t-t')^2 -|{\bf x}_i(t)- {\bf x}_i(t') +{\bf y'}-{\bf y}|^2} \,.
                                       \label{eq:Wvac}
 \end{equation}
The effect of the finite size of the wavepackets is to render this integral
finite. Some explicit examples are discussed in the next section.

     The range of validity of the assumption that the wavepackets do not spread
may be easily estimated. Let us assume that our initial wavepacket is close
to a minimum uncertainty wavepacket, so that the initial size $\Delta x_0$
and the spread in momentum, $\Delta p$, in any direction are related by
\begin{equation}
\Delta x_0 \, \Delta p \approx \frac{1}{2}\,.
\end{equation}
Let $v$ be the mean speed of the wavepacket, and $L$ be the linear distance
to be travelled. The uncertainty in the wavepacket's speed is $\Delta v =
\Delta p/m \approx (2 m \Delta x_0)^{-1}$. During the flight time $L/v$, this
leads to an increase in the size of the packet of the order of 
$L(2 m \Delta x_0 v)^{-1}$. So long as this size increase is small compared 
to the original size $\Delta x_0$, spreading can be ignored. This condition
may be cast as an upper bound on the distance $L$:
\begin{equation}
L \ll 2 \sqrt{2m E}\, (\Delta x_0)^2 \approx 
1{\rm m}\, \biggl(\frac{E}{10\, {\rm keV}}\biggr)^\frac{1}{2} 
\biggl(\frac{\Delta x_0}{ 1 \mu m}\biggr)^2 \,,
\end{equation}
where $E$ is the mean kinetic energy of the electrons. This condition may
be easily satisfied in a realistic experiment. The T\"ubingen group has
performed a number of experiments involving manipulation of wavepackets.
(See Nicklaus and Hasselbach \cite{NH93} and references therein.) In these
experiments, the typical electron energy is of the order of a few keV, and the 
``coherence length'' $\Delta x_0$ varies between 10 nm and 1 $\mu$m.

\section{Particular Trajectories}

\subsection{Parallel lines}

Let us consider the case where the electrons move upon straight paths
of equal length and the wavepackets for both paths are of the same form. 
Let $T$ be the time of flight of an electron upon either
path $C_1$ or $C_2$. The integrals appearing in $W_V$ may be evaluated 
in the electron's rest frame, and we may write the contribution from $C_1$
and $C_2$ together as
\begin{equation}
W_V = -\frac{\alpha}{ \pi} \int {\rm d}^3 {\bf y}\, f({\bf y}) \;
          \int {\rm d}^3 {\bf y}'\, f({\bf y}') \; K(\rho)   \,. 
                              \label{eq:wv}
\end{equation}
Here 
\begin{equation}
K(\rho) =  \int_0^T {\rm d}t\, \int_0^T {\rm d}t'  \, \,
         \frac{1}{(t-t')^2 - \rho^2}
       = 2 \int_0^T {\rm d}\tau \, \frac{T -\tau}{\tau^2 -\rho^2}   \,
                                                         \label{eq:defI3} 
\end{equation}
where $\rho = |{\bf y} - {\bf y}'|$. 
 
The integral in Eq.~(\ref{eq:defI3}) may be evaluated to yield
\begin{equation}
K(\rho) = \frac{T}{\rho}\, \ln \biggl(\frac{T-\rho}{\tau+\rho} \biggr)
          - \ln \biggl(\frac{T^2-\rho^2}{\rho^2} \biggr) \,.
   \label{eq:Iexact} 
\end{equation}
In the limit that $T \gg \rho$, we have
\begin{equation}
K(\rho) \approx -2 - \ln \biggl(\frac{T^2}{\rho^2} \biggr) \,.
   \label{eq:Iapprox} 
\end{equation}
If we combine Eqs.~(\ref{eq:wv}) and ~(\ref{eq:Iapprox}), we obtain
\begin{equation}
W_V = \frac{\alpha}{\pi}\bigg[ 2 - \kappa +
                 2 \ln \biggl(\frac{T}{\ell} \biggr)\biggr] \,, 
                              \label{eq:wv1}
\end{equation}
where 
\begin{equation}
\kappa \equiv  \int {\rm d}^3 {\bf y}\, f({\bf y}) \;
\int {\rm d}^3 {\bf y}'\, f({\bf y}') \;\ln \biggl(\frac{\rho}{\ell} \biggr)\,, 
                              \label{eq:defkappa}
\end{equation}
and $\ell$ is a characteristic length scale associated with the wavepacket.
Note that $W_V$ is independent of $\ell$. We expect that the dimensionless
constant $\kappa$ will be of order unity. Detailed calculations of $\kappa$
for different shapes of the wavepacket will be given in 
Sect.~\ref{sec:kappa}. 

\begin{figure}
\begin{center}
\leavevmode\epsfysize=8cm\epsffile{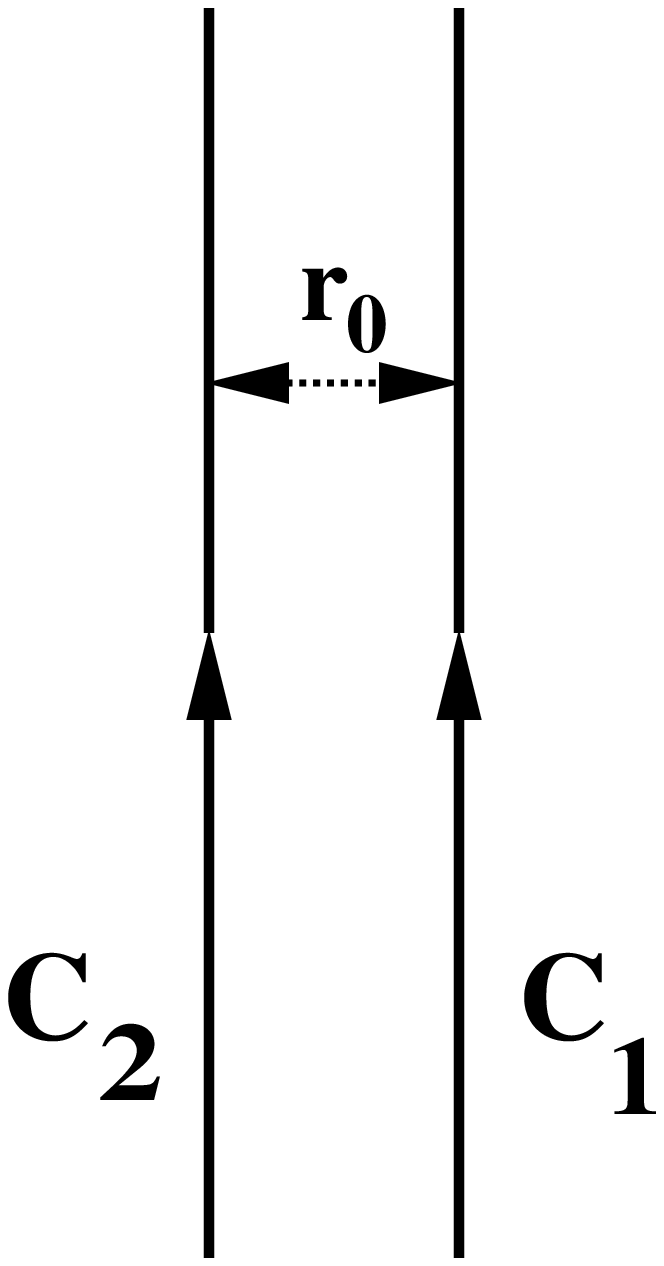}
\label{Figure 1}
\end{center}
\begin{caption}[]

Electrons moving along spacetime paths $C_1$ and $C_2$
travel in space along parallel lines separated by a distance $r_0$. 
\end{caption}
\end{figure}

As discussed above, the photon emission contribution, $W_\gamma$, is
finite even in the limit of classical trajectories. Consequently, if the 
wavepackets are sufficiently localized, we may use Eq.~(\ref{eq:defwg})
to compute $W_\gamma$. Let us consider the specific geometry
illustrated in Fig. 1, where the electrons
travel at constant velocity for a time $T$, with the paths separated by a 
distance $r_0$.  Let $C_1$ be described by the spatial trajectory
${\bf x} = {\bf x}_1(t) = (\frac{1}{2} r_0, vt,0)$ and $C_2$ be described
by ${\bf x} = {\bf x}_2(t') = (-\frac{1}{2} r_0, vt',0)$, where $v$ is the 
electron's speed. Equation~(\ref{eq:defwg}) now becomes
\begin{equation}
W_\gamma = \frac{\alpha}{\pi} \int_0^T {\rm d}t\, \int_0^T {\rm d}t' \,\, 
\frac{1-v^2}{(1-v^2)(t-t')^2 - r_0^2} \approx \, 
\frac{\alpha}{\pi}\, K(r_0)  \,. 
                              \label{eq:wgamma}
\end{equation}
In the limit that $T \gg r_0$, Eq.~(\ref{eq:Iapprox}) yields
\begin{equation}
W_\gamma \approx -2 \frac{\alpha}{ \pi}\,
      \bigg[ 1 + \ln \biggl(\frac{T}{r_0} \biggr)\biggr] \,. 
                              \label{eq:wgamma3}
\end{equation}
If we combine Eqs.~(\ref{eq:wv1}) and ~(\ref{eq:wgamma3}), the result
is
\begin{equation}
W = W_V + W_\gamma = \frac{\alpha}{\pi}\,
      \bigg[ 2\,\ln \biggl(\frac{r_0}{\ell} \biggr) -\kappa \biggr] \,. 
                              \label{eq:wfinal}
\end{equation}
 Note that although the magnitudes of both $W_V$ and $W_\gamma$ grow 
logarithmically in time, $W$ is constant in the limit that $T \gg r_0$.

\subsection{An example of intersecting paths}
\label{sec:int}

Although the example of the previous subsection is useful for understanding
the interplay between $W_V$ and $W_\gamma$, it is not a realistic description
of what might actually occur in an electron interferometer. The electrons
emerge from a single coherent source, are separated into two distinct paths, 
and are later recombined to form the interference pattern. ( For a review
of the construction of electron interferometers, see Missiroli {\it et al}
\cite{MPV81}.) Thus a somewhat
more realistic pair a classical paths is that illustrated in Fig. 2.
We will assume that the wavepackets move with a group velocity whose magnitude
$v$ is constant along each segment of the trajectory. We also assume that
$L_2 \gg L_1$, so the longest period of time is spent on the parallel straight
line segments, $b$. 

\begin{figure}
\begin{center}
\leavevmode\epsfysize=12cm\epsffile{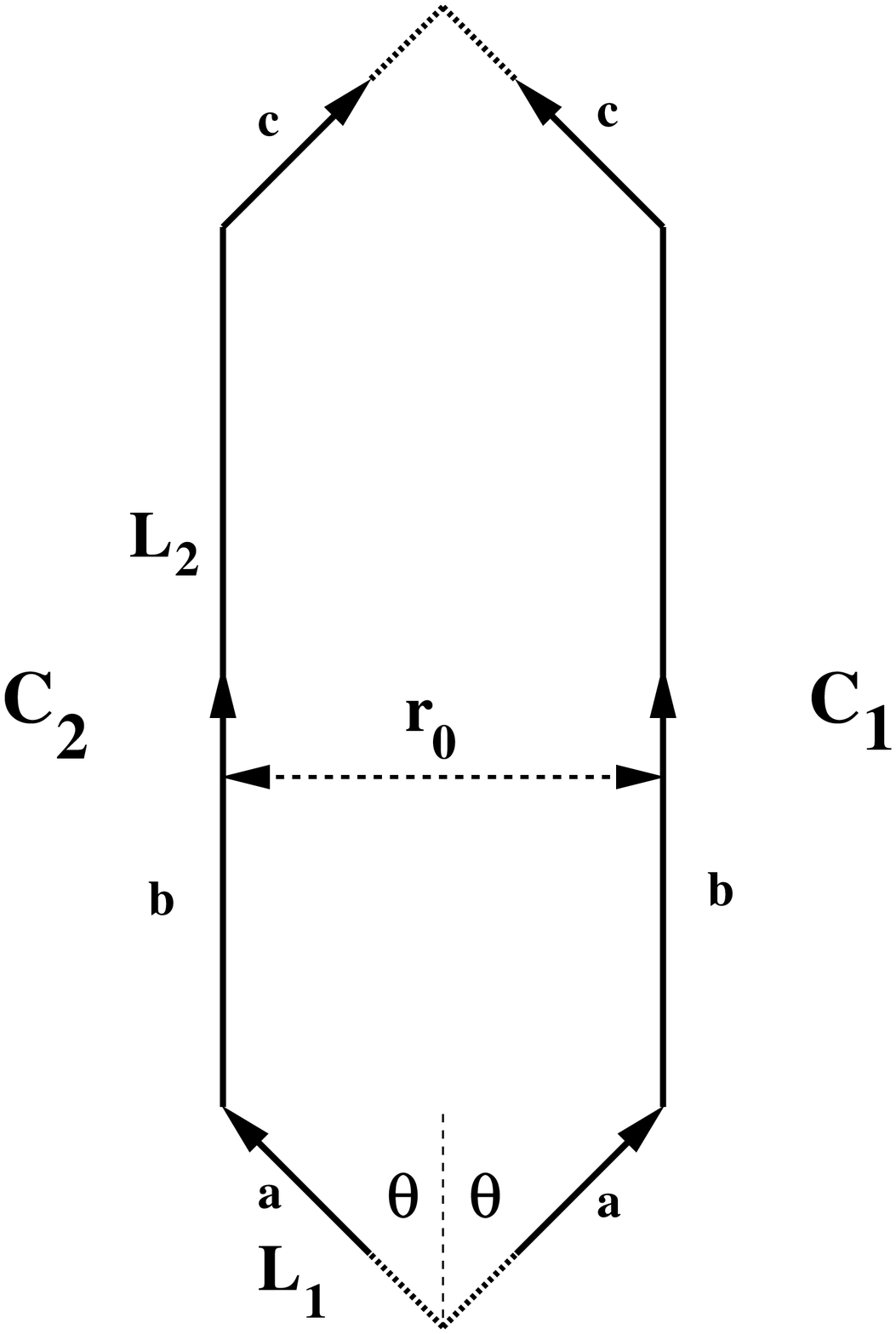}
\label{Figure 2}
\end{center}
\begin{caption}[]

The spatial geometry of a pair of intersecting paths 
is illustrated. The dotted lines near the lower and upper apexes denote
the fact that these classical trajectories are only applicable when the
distance from the apex is greater than the longitudinal wavepacket size, 
$\ell$. 
\end{caption}
\end{figure}

\subsubsection{Vacuum Part}

 We may write Eq.~(\ref{eq:Wvac}) as 
\begin{equation}
W_V = - \frac{\alpha}{2 \pi}\, J \, , \label{eq:defJ}
\end{equation}
where the integral $J$ may be evaluated along either of $C_1$ or $C_2$, and
can be expressed as a sum of terms:
\begin{equation}
J = J_{aa} + J_{bb} +J_{cc} +J_{ab} +J_{ba} +J_{ac} +J_{ca} +J_{bc} +J_{cb}\,,
                                                             \label{eq:Jsum1}
\end{equation}
where $J_{ij}$ arises when ${\bf x}$ is on segment $i$ and ${\bf x}'$ is on 
segment $j$. By symmetry, we have $J_{ab} = J_{ba}$, $J_{bc} =J_{cb}$, and
$J_{ac} =J_{ca}$ . Furthermore, when wavepacket spreading is ignored, 
$J_{aa} \approx J_{cc}$ and $J_{ab} \approx J_{bc}$. Finally, in the limit
that $L_2 \gg L_1$, we may ignore $J_{ac} =J_{ca}$ compared to the other terms.
Thus we have
\begin{equation}
J \approx 2J_{aa} + J_{bb} + 4J_{ab} \,. \label{eq:Jsum2}
\end{equation}
The contributions $J_{aa}$ and $J_{bb}$, where both ${\bf x}$ and ${\bf x}'$
lie on the same straight line segment, were calculated above. 
  From Eq.~(\ref{eq:wv1}), we have, in the limit that $v \ll 1$,
\begin{equation}
J_{aa} \approx  -2 + \kappa - 2 \ln \biggl(\frac{L_1}{\ell v} \biggr)
                                                          \label{eq:Jaa}
\end{equation}
and 
\begin{equation}
J_{bb} \approx  -2 + \kappa - 2 \ln \biggl(\frac{L_2}{\ell v} \biggr) \,.
                                                          \label{eq:Jbb}
\end{equation}
In the integral $J_{ab}$, we may take ${\bf x}$ to lie on segment $a$ and 
${\bf x}'$ to lie on $b$. Because the size of the wavepackets is small
compared to the lengths of these segments ($\ell \ll L_1 \ll L_2$), we
may use the approximation of classical trajectories, except near the 
intersection point of the two segments. Segment $a$ is described by 
\begin{equation}
x = v t \sin\theta, \quad y = v t \cos\theta , \qquad 0\leq t \leq T_1 \,,
                                                  \label{eq:apath}
\end{equation}
and segment $b$ by
\begin{equation}
x' = v T_1 \sin\theta, \quad 
y' = v T_1 \cos\theta + v(t' -T_1), \qquad T_1 \leq t' \leq T_2 \,.
\end{equation}
To avoid a divergence at $t = t' =T_1$, we change the ranges of integration
to be $0\leq t \leq T_1 - \tau/2$ and $T_1 + \tau/2 \leq t' \leq T_2$, where
$\tau = \ell/v$ is the characteristic time required for the wavepacket to
pass a given point. We now have
\begin{eqnarray}
&&  \!\!\!\!\!\!\!\!\!  J_{ab} \approx   \nonumber \\
&& \!\!\!\!\!\!\!\!\!  \int_0^{T_1 -\frac{1}{2}\tau} dt 
                 \int_{T_1 + \frac{1}{2}\tau}^{T_2} d t'
\; \frac{1 -v^2 \cos^2 \theta}
{(t-t')^2 -v^2 \sin^2 \theta \, (t-T_1)^2 
      -v^2[t\cos\theta -T_1 \cos\theta +(T_1-t')]^2} \nonumber \\
&\approx& \int_0^{T_1 - \frac{1}{2}\tau} dt 
            \int_{T_1 + \frac{1}{2}\tau}^{T_2} d t'
\; \frac{1}{(t-t')^2} + O(v^2) 
= \ln \left[\frac{(T_1 +\frac{1}{2}\tau)(T_2 +\frac{1}{2}\tau)}
                 {\tau (T_2 + T_1)} \right] \,.
\end{eqnarray}
Use $L_1=T_1v$, $L_2=T_2v$, and $\ell \ll L_1 \ll L_2$ to obtain the result
\begin{equation}
J_{ab} \approx \ln \biggl(\frac{L_1}{\ell} \biggr) \,.  \label{eq:Jab}
\end{equation}
We may now combine Eqs.~(\ref{eq:defJ}), (\ref{eq:Jsum2}), (\ref{eq:Jaa}),
(\ref{eq:Jbb}), and (\ref{eq:Jab}) to obtain our result for the vacuum
fluctuation term for this trajectory
\begin{equation}
W_V = \frac{\alpha}{2 \pi}\bigg[ 3(2 - \kappa) +
                 2 \ln \biggl(\frac{L_2}{\ell v^3} \biggr)\biggr] \,. 
                              \label{eq:wv2}
\end{equation}
Note that the dependence upon the length of the longer path, $L_1$,
has cancelled out.  

\subsubsection{Photon Emission Part}

  Now we wish to calculate $W_\gamma$ for the paths shown in Fig. 2.
In a notation analogous to that used in Eq.~(\ref{eq:Jsum1}), we may write
\begin{equation}
W_\gamma = \frac{\alpha}{2 \pi}\, I \, , \label{eq:defI2}
\end{equation}
and
\begin{equation}
I = I_{aa} + I_{bb} +I_{cc} +I_{ab} +I_{ba} +I_{ac} +I_{ca} +I_{bc} +I_{cb}\,,
                                                             \label{eq:Isum1}
\end{equation}
The same symmetry considerations and approximations used to obtain
Eq.~(\ref{eq:Jsum2}) now lead to 
\begin{equation}
I \approx 2I_{aa} + I_{bb} + 4I_{ab} \,. \label{eq:Isum2}
\end{equation}
The contribution of the two parallel segments, $I_{bb}$, was computed 
in Eq.~(\ref{eq:wgamma3}), and can be expressed as
\begin{equation}
I_{bb} \approx 
 -2 \bigg[ 1 + \ln \biggl(\frac{L_2}{2L_1 v \sin \theta} \biggr)\biggr] \,. 
                              \label{eq:Ibb}
\end{equation}

    The integral $I_{aa}$ would diverge if we allow the points ${\bf x}$
and ${\bf x'}$ to coincide at the vertex. However, the classical trajectory
approximation is not valid all the way to this point. A more realistic
approximation is to cut off the integrations a finite distance from the
vertex of the order of the wavepacket size $\ell$. Thus, we take the 
integrations in $I_{aa}$ to lie in the range $\ell/v \leq t, t' \leq T_1$
and write
\begin{equation}
I_{aa} = \int_{\frac{\ell}{v}}^{T_1} dt \, \int_{\frac{\ell}{v}}^{T_1} d t'
\; \frac{1 -v^2 (\cos^2 \theta - \sin^2 \theta )}
{(t-t')^2 (1 - v^2 \cos^2 \theta) -v^2 \sin^2 \theta \, (t + t')^2 } \,.
\end{equation}
We may ignore the $v^2$ terms in the numerator and in the $(t-t')^2$ term
in the denominator, and write
\begin{equation}
I_{aa} \approx \int_{\frac{\ell}{v}}^{T_1} dt\, \int_{\frac{\ell}{v}}^{T_1} d t'
\; \frac{1}{(t-t')^2 -v^2 \sin^2 \theta \, (t + t')^2 } \,.
\end{equation}
We cannot, however, ignore the remaining $v^2$ dependence in the denominator,
as this would result in a divergent integral.
This integral may be evaluated using the symbolic manipulation routine
MACSYMA, and then expanded for small $v$, with $\ell/v$ fixed, 
to yield the result
\begin{equation}
I_{aa} \approx \ln \biggl( \frac{\ell v^2 \sin^2 \theta}{L_1} \biggr) 
               + 2(\ln 2 -1) \,.
\end{equation}

    Finally, we have the integral $I_{ab}$, in which we may take ${\bf x}$
to lie on segment $a$ of $C_1$ and be described by Eq~(\ref{eq:apath}), and
take ${\bf x'}$ to lie on segment $b$ of $C_2$ and be described by
\begin{equation}
x' = - v T_1 \sin\theta, \quad 
y' = v T_1 \cos\theta + v(t' -T_1), \qquad T_1 \leq t' \leq T_2 \,.
\end{equation}
We can express $I_{ab}$ as
\begin{eqnarray}
&&  \!\!\!\!\!  I_{ab} = \nonumber \\
&& \!\!\!\!\!  \int_0^{T_1} dt     \int_{T_1}^{T_2 + T_1} d t'
\; \frac{1 -v^2 \cos^2 \theta}
{(t-t')^2 -v^2 \sin^2 \theta \, (t+ T_1)^2 
      -v^2[t\cos\theta -T_1 \cos\theta +(T_1-t')]^2} \approx \nonumber \\
&&\!\!\!\!\! \int_0^{T_1} dt     \int_{T_1}^{T_2 + T_1} d t'
\; \frac{1} {(t-t')^2 -v^2 \sin^2 \theta \, (t+ T_1)^2 
      -v^2[t\cos\theta -T_1 \cos\theta +(T_1-t')]^2}  \nonumber \\
&& + \,O(v^2)\,.
\end{eqnarray}
That is, we may ignore the $v^2$ term in the numerator, but not those in the
denominator, as this would result in an integral which is divergent at
$t=t'=T_1$. However, away from this point, the $v^2$ terms in the denominator
are negligible, and we may set $t=t'=T_1$ in these terms, and write
\begin{equation}
I_{ab} \approx \int_0^{T_1} dt     \int_{T_1}^{T_2 + T_1} d t'
\; \frac{1} {(t-t')^2 - 4 T_1^2 v^2 \sin^2 \theta } \,.
\end{equation}      
This integral may be evaluated in terms of logarithm functions (e.g. by
use of MACSYMA). In the limit that $v \ll 1$ and $T_2 \gg T_1$, the result
becomes
\begin{equation}
I_{ab} \approx 1 - \ln (2 v \sin\theta) + O(v \ln v) \,.
\end{equation}

     We may now combine these results and write
\begin{equation}
W_\gamma = - \frac{\alpha}{\pi}\bigg[ 1 - \ln 2  +
          \ln \biggl(\frac{L_2}{\ell v \sin \theta} \biggr)\biggr] \,, 
                              \label{eq:wgamma4}
\end{equation}
and hence
\begin{equation}
W = W_V + W_\gamma = \frac{\alpha}{2 \pi}\,
 \bigg[ 2\,\ln \biggl(\frac{2 \sin\theta}{v^2} \biggr) +4 -3\kappa \biggr] \,. 
                              \label{eq:wfinal2}
\end{equation}
Note that the dependences upon both the length $L_2$ and the wavepacket
size $\ell$ have cancelled out in this result.

\section{Calculation of $\kappa$ for Particular Wavepackets}
\label{sec:kappa}  

\subsection{Spherical Wavepackets}

    Let us here consider a wavepacket whose probability density is constant
within a sphere of radius $R$, and zero otherwise: $f({\bf x}) = f(r)
= 3/(4\pi R^3)$ for $r \le R$, and $f = 0$ for $r > R$. Let us take the
characteristic length $\ell$ in this case to be the diameter of the sphere,
$\ell = 2R$. Then the integrals appearing in Eq.~(\ref{eq:defkappa}) may
be explicitly evaluated to yield
\begin{equation}
\kappa_{\rm sphere} = - \frac{3}{2}\,.
\end{equation}

\subsection{Cylindrical Wavepackets}

    Now we wish to consider a more general class of wavepackets which
are characterized by more than one length scale. An example is a 
wavepacket whose probability density is constant within a cylinder of 
radius $R$ and length $L$, so that in a cylindrical coordinate system
\begin{equation}
f = \left\{ \begin{array}{ll}
 {\pi R^2 L}^{-1} & \mbox{if $r\le R$ and $0 \le z\le L$} \nonumber \\ 
         0        & \mbox{otherwise.}
          \end{array}
\right.
\end{equation}
In this case, three of the six integrations appearing in 
Eq.~(\ref{eq:defkappa}) may be performed explicitly, with the result
\begin{equation}
\kappa = \frac{4}{\pi}\, \int_0^1 {\rm d}\rho \,\rho 
\int_0^1 {\rm d}\rho' \,\rho' \int_0^{2\pi} {\rm d}\phi \, 
                                            F(\rho,\rho',\phi) \,.
                              \label{eq:kappacyl}
\end{equation}
Here 
\begin{equation}
F(\rho,\rho',\phi) = \ln \biggl(\frac{R}{\ell} \biggr) +
\beta^{-2} \biggl\{ b^2 \ln b - \frac{1}{2} \Bigl[ 
  (b^2 +\beta^2) \ln (b^2 +\beta^2) - 
4\beta b \arctan\Bigl(\frac{\beta}{b}\Bigr) +3\beta^2\biggr]\biggl\}\,,
\end{equation}
with $\beta = L/R$ and $b =\sqrt{\rho^2 -2\rho\rho'\cos \phi +\rho'^2}$.

Let us first examine the limit in which $\beta \rightarrow 0$ ($L \ll R$).
In this case,
\begin{equation}
F \sim \ln \biggl(\frac{R}{\ell} \biggr) + \ln b +O(\alpha^2) \,. 
\end{equation}
If we set $\ell =2R$, the diameter of the cylinder, then we can see that
$\kappa$ will be of order unity in this case ($\ln b \sim O(1)$ except
in regions where $|b| \ll 1$, which yield a negligible contribution to
the total integral).

Now let us examine the limit in which $\alpha \rightarrow \infty$ 
($L \gg R$), in which case
\begin{equation}
F \sim - \ln \biggl(\frac{L}{\ell} \biggr) +O(1) \,. 
\end{equation}
Now we will obtain  $\kappa$ of order unity if we set $\ell = L$.
At least for the case of a cylinder, in order to keep $\kappa$ of order 
unity, we should always take $\ell$ to be the {\it larger} of the two 
length scales which characterize the wavepacket. The results of a 
numerical evaluation of $\kappa$ are illustrated in Fig. 3. The cusp arises
from the fact that $\ell = L$ when $L > 2R$, and $\ell =2R$ when $L < 2R$.

\begin{figure}
\begin{center}
\leavevmode\epsfysize=8cm\epsffile{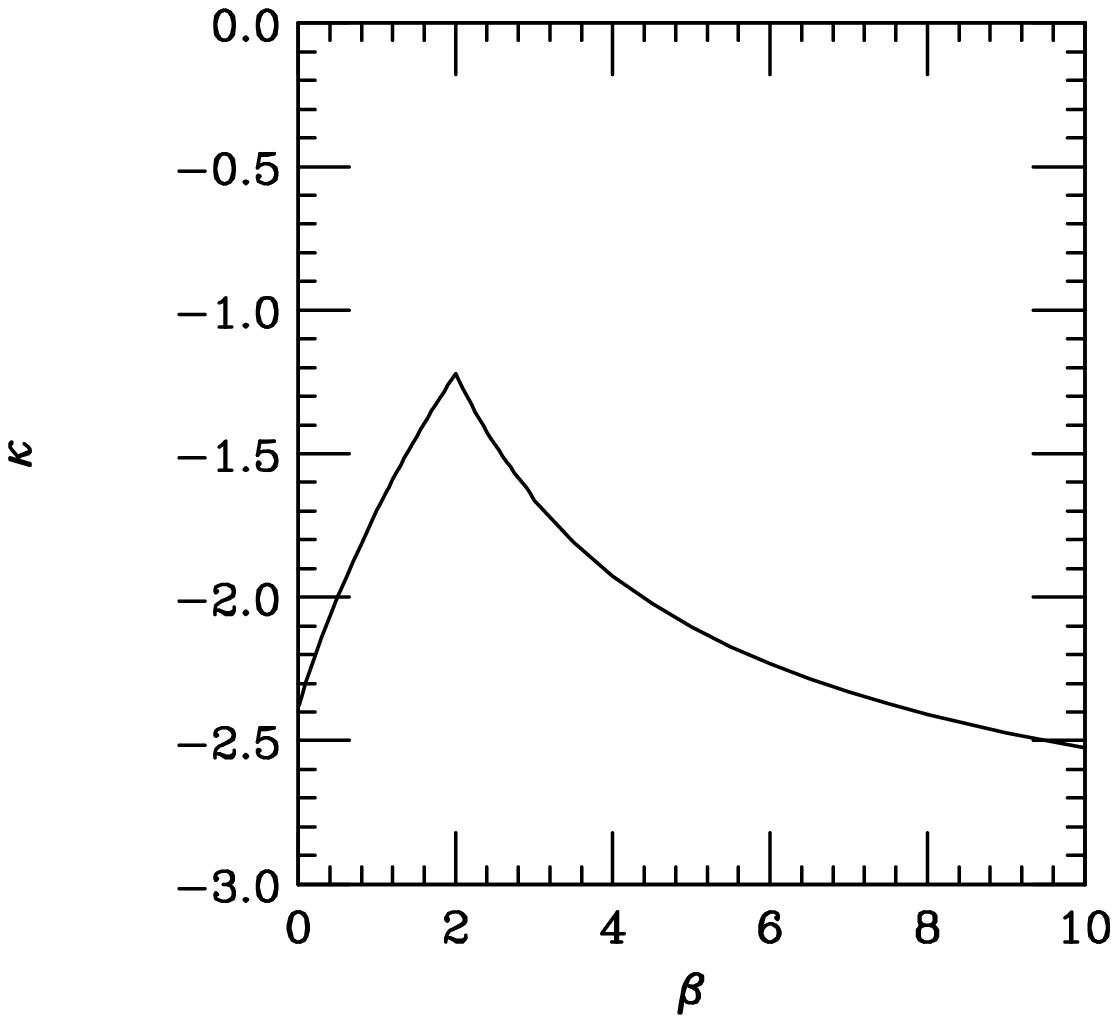}
\label{Figure 3}
\end{center}
\begin{caption}[]

The parameter $\kappa$ is plotted as a function of $\beta =
L/R$ for cylindrical wavepackets of length $L$ and radius $R$.
\end{caption}
\end{figure}

\section{Summary and Conclusions}
\label{sec:sum}

   In the previous sections, we have found that the finite sizes of electron
wavepackets do indeed remove the ultraviolet divergence which arises
when one attempts to evaluate the vacuum fluctuation effects on an interference
pattern, $W_V$, in the approximation of classical
trajectories, Eq.~(\ref{eq:defwv}). The resulting expressions for $W_V$,
e.g. Eq.~(\ref{eq:wv1}), now acquire a logarithmic dependence upon $\ell$, the
maximum spatial extent of the wavepacket. However, in the geometry discussed
in Sec.~\ref{sec:int}, illustrated in Fig. 2, the photon emission term, 
$W_\gamma$, acquires a compensating dependence. This arises from the fact 
that the classical trajectory approximation fails when one is within a distance 
$\ell$ of the beginning and ending point of the trajectories. As a result,
the final result for this geometry, Eq.~(\ref{eq:wfinal2}), is independent of
$\ell$. 

In principle, it is possible to observe the effects of vacuum fluctuations
and of photon emission separately. As discussed in I, this might be done
in a veto experiment in which one counts only those electrons which have not 
emitted photons, and hence measures  $W_V$ alone. In practice, this would 
be extremely difficult to implement, as the photon energies involved are
likely to be very low. Nonetheless, $W_V$ and $W_\gamma$ are each physically
meaningful quantities. This brings us to the question of understanding the 
signs of each quantity separately. From Eqs.~(\ref{eq:wgamma3}) or
(\ref{eq:wgamma4}), we see that typically $W_\gamma < 0$. This result is 
readily understood as the decohering effect of photon emission. From  
Eq.~(\ref{eq:numdens1}), we see that $W <0$ means a decrease in the
amplitude of the interference oscillations. A photon whose wavelength is
less than the separation between the two electron paths carries information
about which path the electron has taken, and hence naturally has a decohering
effect. 

   The sign of $W_V$ is somewhat more problematic. We can see from
Eqs.~(\ref{eq:wv1}) or ~(\ref{eq:wv2}) that typically $W_V > 0$. Indeed, 
as $\ell \rightarrow 0$, we obtain the divergent results of the classical
trajectory approximation, $W_V \rightarrow + \infty$. One can never turn
off the effects of the vacuum fluctuations, but only modify them. One
can interpret the dependence of $W_V$ upon $\ell$ as indicating that the
decohering effects of vacuum fluctuations are suppressed for smaller
wavepackets as compared to larger wavepackets. In any case, $W_V$ can 
never become sufficiently large as to cause $n < 0$.

    Note that the final result of Sect. \ref{sec:int}, Eq.~(\ref{eq:wfinal2}),
is a positive value of $W$ which increases slowly as either $\sin\theta$
or $v$ decrease. The net change in the amplitude of the interference 
oscillations is typically of the order of $1 \%$. In principle this effect
is observable, although its detection may be difficult.  
 The best hope for observing
the effects of vacuum fluctuations upon electron coherence seems to lie in
experiments in which one looks at shifts in the amplitude of the interference
oscillations due to changes in the photon two point function \cite{F93,F95}.
The results of the present paper indicate that the approximation
of classical trajectories is remarkably robust, and may be applied to the 
analysis of such experiments.

\vspace{0.5cm}
{\bf Acknowledgements:} I would like to thank E. Yablonovich for a 
discussion which inspired this paper. I would also like to thank
F. Hasselbach, G. Matteucci, and G. Pozzi for helpful discussions on
the experimental aspects of electron interferometry. This work was
supported in part by the National Science Foundation under Grant
PHY-9507351.


\vspace{1cm}

\begin{thebibliography}{--}

\bibitem{F93} L.H. Ford, Phys. Rev. D {\bf 47}, 5571 (1993).
\bibitem{units} Except as otherwise noted, units in which $c = \hbar =1$
will be used. Electromagnetic quantities are in Lorentz-Heaviside units, so
the fine structure constant is $\alpha = e^2/4\pi$, where $e$ is the electron
charge.

\bibitem{AB} Y. Aharonov and D. Bohm, Phys. Rev. {\bf 115}, 485 (1959).

\bibitem{AS} A. Stern, Y. Aharonov and Y.Imry, Phys. Rev. A {\bf 41},
             3436 (1990).

\bibitem{F95} L.H. Ford, Ann. N. Y. Acad. Sci. {\bf 755}, 741 (1995).

\bibitem{Schwinger} J. Schwinger, Phys. Rev. {\bf 152}, 1219 (1966);
{\bf 158} (1967).

\bibitem{NH93} M. Nicklaus and F. Hasselbach, Phys. Rev. A {\bf 46}, 152 (1993).

\bibitem{MPV81} G.F. Missiroli, G. Pozzi and U. Valdr{\`e}, J. Phys. E
                {\bf 14}, 649 (1981).

\end{thebibliography}
\end{document}